\documentclass[pra,twocolumn,preprintnumbers,amsmath,amssymb,superscriptaddress,showpacs,longbibliography]{revtex4-2}
\usepackage{graphicx}
\usepackage{latexsym}
\usepackage{amsmath}
\usepackage{graphics}
\usepackage{amssymb}
\usepackage{layout}
\usepackage{verbatim}
\usepackage{amsfonts,epsfig,dsfont}
\usepackage{natbib}
\usepackage{hyperref}
\usepackage{bbold}
\usepackage{enumerate}
\usepackage{tabularx}
\usepackage{floatrow}
\newcommand{\ket}[2][]{{|#2\rangle_{#1}}}
\newcommand{\bra}[2][]{{}_{#1}\langle #2|}

\def\duzomniejsze{<\kern-.7mm<}
\def\duzowieksze{>\kern-.7mm>}

\def\textbf#1{{\bf #1}}
\def\beq{\begin{equation}}
\def\eeq{\end{equation}}
\def\be{\begin{equation}}
\def\ee{\end{equation}}
\def\ben{\begin{eqnarray}}
\def\een{\end{eqnarray}}
\def\beqa{\begin{eqnarray}}
\def\eeqa{\end{eqnarray}}
\def\eea{\end{array}}
\def\bea{\begin{array}}

\usepackage{color}
\definecolor{dgreen}{RGB}{0,90,0}

\begin{document}

\title{Estimation of nuclear polarization via discrete measurement of NV center spin evolution}

\author{Mateusz Kuniej}
\affiliation{Institute of Theoretical Physics, Faculty of Fundamental Problems of 
	Technology, Wroc{\l}aw University of Science and Technology,
	50-370 Wroc{\l}aw, Poland}
\author{Katarzyna Roszak}
\affiliation{FZU - Institute of Physics of the Czech Academy of Sciences, 182 00 Prague, Czech Republic}

\date{\today}

	\begin{abstract}
    We propose a method for the estimation of the initial polarization of spinful nuclei of the ${}^{13}C$ isotope
    in diamond via a measurement of the evolution of the coherence of an NV center spin qubit. 
    Existing polarization measurement methods are difficult to implement experimentally, because they require direct interference in the environment of the qubit. Here, in
    order to obtain the 
    information, it is necessary to measure the qubit coherence at certain points of time, which are unambiguously 
    determined by the applied magnetic field. For sufficiently high magnetic fields, the minimum value of
    the measured coherence constitutes an upper bound on the product of the initial polarizations of each environmental
    spin. The most significant advantage of the method, which allows to infer initial values of nuclear 
    polarizations without any direct access to the environment, lies in its simplicity and the small amount
    of experimental resources that it requires. We exemplify the operation of the scheme on a realistic, randomly
    generated environment of eight nuclear spins, obtaining a reasonably accurate estimation of the initial polarization.
	\end{abstract}

	\maketitle
	
	
	\section{Introduction \label{sec1}}
	
	Nitrogen-vacancy (NV) centers in diamond \cite{doherty13,wood18,awschalom18,tchebotareva19} are attracting a lot of attention for 
	their applications in ultrasensitive magnetometry \cite{maze08b, Taylor_NTRP_08,Hall_PRL_09, delange11,Hall_MRS_13,sasaki17,sasaki18}.
	Since these applications rely on the effect that even very small magnetic fields have
	on the evolution of the NV center spin qubit, they
	require minimization of decoherence stemming from sources within the diamond lattice
	where the NV center is embedded.
	The dominating decoherence mechanism here is the hyperfine interaction with spinfull ${}^{13}C$
	carbon atoms present in the lattice \cite{zhao12,kwiatkowski18}.
	Since the ${}^{13}C$ isotope is rare (and randomly distributed within the lattice),
	the NV center spin interacts with a small number of environmental spins \cite{maze08,zhao11},
	which leads to naturally occurring long decoherence times \cite{kennedy03,gaebel06,delange11}.
	Nevertheless, even slower decoherence is necessary for effective ultrasensitive magnetometry,
	so techniques such as dynamical decoupling (DD) \cite{viola99,delange10,ryan10,naydenov11} and dynamical nuclear polarization (DNP) \cite{london13,fischer13,pagliero18,wunderlich17, scheuer17,poggiali17,hovav18,henshaw19}
	have to be employed. 

    Recently, several schemes for polarizing nuclear spins and its detection have been developed.
    A very promising DNP method is based on optically pumped NV centers where several experimental demonstrations have been reported \cite{Jacques_PRL_2009, london13, fischer13, King_NTCOM_2015, Alvarez_NTCOM_2015}. The main disadvantage of those schemes is their sensitivity to the magnetic field alignment along the NV crystal axis. 
    The Polarization readout by Polarization Inversion (PROPI) technique allowed to find new polarization schemes that can be applied at arbitrary magnetic field strengths and a wide range of magnetic field orientations \cite{scheuer17}.
    
    One of the detection schemes reads out the magnetization of the nuclear spin bath using PROPI, which provides a measure of polarization of the nuclear spins surrounding the NV center and allows observation of hidden polarization dynamics similar to Landau-Zener-St{\"u}ckelberg oscillations \cite{scheuer17}. More often estimation of the value of the polarization is achieved using optically detected magnetic resonance (ODMR) by integrating electron resonances \cite{Jacques_PRL_2009} or characterizing the ODMR peaks by transition probabilities in the hyperfine basis states \cite{Busaite_PRB_2020}. Another intriguing scheme exploits the Fourier-transformed nuclear magnetic resonance spectrum in which Lorentzian peaks are proportional to the nuclear polarization \cite{King_NTCOM_2015}.

    In general, measuring the polarization of nuclear spins is challenging due to their miniature magnetic moment. One of the possible solution is to use large ensemble of them \cite{fischer13} or to search for an energy shift of electronic spin due to the interaction between it and static polarized bath \cite{makhonin11}. The latter method strongly depends on the geometrical configuration of the nuclear spin ensemble.
    
    Here we propose a much simplified scheme for the detection of polarization of the nuclear environment
    which requires only the measurement of the evolution of the coherence of the NV center spin qubit, instead of measurements on the nuclear spins directly. 
    The qubit evolution in principle contains the full information about the initial state of the environment. Yet, since the interaction with each nuclear spin is different, due to the strong dependence on the actual location of the spinful carbon isotope in the diamond lattice, and the strength of the coupling between the qubit with each spin of the environment is unknown, a straightforward readout of the initial polarization from the qubit evolution is not feasible. 
    
    Nevertheless it is possible to determine the joint effect that the spin-bath has
    on the NV-center which is proportional to the sum of individual nuclear spin polarizations weighted by the 
    strengths of the coupling characteristic for each environmental spin. In fact, a method for probing nuclear bath polarization as such a weighted 
    average using a spin-echo based
    technique has been demonstrated \cite{london15}. There the weighted average of the polarization was determined
	from the phase accumulation rate and revival times obtained using quadrature detection. However, this method works better if it is applied to NV ensembles, rather than to the single qubit in contrast to the FID measurements, where diminishing FID signal requires higher signal-to-noise ratio.
    
    The technique proposed in this paper uses a the simplest possible measurement of the coherence of a single NV-center (no echo).
    It relies on the fact that the coherence evolution 
    is dependent on a frequency that is a direct function of 
    the applied homogeneous magnetic field and is the same for all nuclear spins. This allows us to determine time-instants for which the function governing the effect
    of each nucleus on the qubit is greatly simplified. A discrete measurement of qubit coherence, in the sense that only the evolution at points of time
    which are governed by simplified functions is probed, allows to obtain much greater information about 
    the environment than the full coherence evolution does. 
    
    We find that this technique, which is limited to the measurement of qubit coherence at points of time
    that are unambiguously determined by the applied magnetic field, allows to find the upper bound on 
    the product of initial environmental polarizations. The drawbacks of the method include the need for measurements
    up to relatively long times and the fact that the obtained value is only an estimate of the actual polarization.
    On the other hand, the straightforward nature of the technique allows for extracting information
    about the environment without any measurements on it, and only accessing the NV center spin qubit
    in a vary basic way. As such, the method should be useful for preliminary studies of the effectiveness of the 
    methods used to polarize the spin environment.
    
    The article is organized as follows. In Sec.~\ref{sec:sys&H} we discuss the evolution of the spin qubit
    in the presence of a nuclear environment. In Sec.~\ref{sec2} we show how probing the qubit at discrete
    points of time can lead to a simplification of the observed qubit evolution in such a way that allows
    to obtain more information about the environment. We discuss two such regimes in Secs \ref{sec2a} and \ref{sec2b}
    and show exemplary results. Sec.~\ref{sec6} concludes the paper.


	\section{Nuclear environment and spin qubit} \label{sec:sys&H}
	   
   The diamond lattice consists mainly of ${}^{12}C$ spinless carbon atoms, but there is about $1.1\%$ of the ${}^{13}C$ spinful carbon isotope. The latter interact with the NV center spin via the hyperfine interaction resulting in the dominant decoherence mechanism, so they constitute the spin environment \cite{Markham_DRM_11}. 
   Hence, when describing the open-system dynamics of the NV center spin, only a small number of environmental spins need to be considered.
   
   The free Hamiltonian of the spin environment is given by
   \begin{equation}
   \hat{H}_E = \gamma_nB_z\sum_{k=0}^{N-1}\hat{I}_k,
   \label{he}
   \end{equation}
   where the magnetic field $B_z$ is applied in the $z$ direction and $\gamma_n = 10.71$~MHz/T is the gyromagnetic ratio of the ${}^{13}C$ nuclei. $\hat{\bm{I}}_k$ is the spin operator for the $k-$th nuclear spin. A term describing the internuclear magnetic interactions can be omitted since they are weak in comparison with 
   the qubit-environment interaction and do not affect any of the described processes in
   a substantial way \cite{zhao12}.
   
   To probe the initial polarization of the environment, we will employ
   a spin qubit defined on two of the low energy NV center spin states, which
   constitute an effective electronic spin triplet $S = 1$ \cite{XingFei_PRB_93}, with the spin levels $\ket{-1}$, $\ket{0}$, $\ket{1}$ being the eigenvectors of the $\hat{S}_{z}$ operator.
   The most widely employed qubit consists of the $m = 0, 1$ states, and in the qubit subspace the free qubit Hamiltonian is given by
   \begin{equation}
   \label{hq}
   \hat{H}_Q = (\Delta -\gamma_{e}B_{z}) \ket{1}\bra{1}.
   \end{equation}
   Here $\Delta = 2.87$~GHz is the zero-field splitting and $\gamma_e = 28.08$~MHz/T is the electron gyromagnetic ratio. 
   
  The hyperfine interaction \cite{Smeltzer_NJOP_11} between this qubit and environment is described by the
  Hamiltonian (the $m=0$ qubit state is decoupled from the environment)
  \begin{equation}
  \label{hqe}
  \hat{H}_{QE}  =
  \ket{1}\bra{1} \otimes\sum_{k=0}^{N-1}\sum_{j\in(x,y,z)}\!\!\!\mathbb{A}_k^{z,j}\hat{I}_k^j.
  \end{equation}
    We have omitted the Fermi contact interaction which describes the non-zero probability of finding the electron bound to the NV center on the location of a given nucleus and is proportional to the electronic wavefunction density at the location of a nucleus. Since the wavefunction of deep defects is strongly localized, the Fermi contact term gives a measurable contribution to the hyperfine coupling for nuclei that are at most $0.5$~nm away from the defect \cite{Gali_PRB_08}. The coupling constants in Eq.~(\ref{hqe}) are given by
    \begin{equation}
    \label{duzea}
        \mathbb{A}_k^{z,j} = \frac{\mu_0}{4\pi}\frac{\gamma_e\gamma_n}{r_k^3}\left(1 - 3\frac{(\textbf{r}_k\cdot\hat{\textbf{j}})(\textbf{r}_k\cdot\hat{\textbf{z}})}{r_k^2}\right),
    \end{equation}
    where $\mu_0$ is the magnetic permeability of the vacuum and the $\textbf{r}_k$ is a displacement vector between the NV center and the $k$-th spin. $\hat{\textbf{j}}=\hat{\textbf{x}},\hat{\textbf{y}},\hat{\textbf{z}}$
    are versors in three distinct directions.
   
   Since the qubit Hamiltonian (\ref{hq}) and the interaction term (\ref{hqe}) commute, the evolution of the qubit is limited to pure decoherence, a process during which only the 
   off-diagonal elements of the qubit density matrix (written in the 
   $\{|0\rangle,|1\rangle\}$ pointer basis of the qubit) evolve and decay.
   In the following, we consider the initial state of the qubit always to be the equal superposition
   state $\ket{\Psi} = \frac{1}{\sqrt{2}}(\ket{0} + \ket{1})$.
   
   The initial state of the spinful carbon nuclei ${}^{13}C$ after DNP is given by
   \begin{equation}
   \label{inibig}
   \hat{R}(0) = \bigotimes_{k} \hat{\rho}_{k}(0),
   \end{equation} 
   where we assumed that there are no initial correlations between the nuclei, and
   $\hat{\rho}_{k}(0)$ denotes the density matrix of $k$-th nucleus (with spin-$1/2$ nuclei) given by
   \begin{equation}
   \label{ini}
   \hat{\rho}_k(0) =\frac{1}{2}(\mathds{1} + 2p_{k}\hat{I}^{z}_{k}).
   \end{equation} 
   Here $p_k \! \in \! [-1,1]$ is the polarization of the $k$-th nucleus. Without DNP the polarization $p_k$ is zero for all nuclear spins, and the initial state of the environment is fully mixed. 
   
   Since there are no inter-nuclear correlations in the environmental initial state and no terms in the Hamiltonian that can induce such correlations, the evolution of the qubit coherence can be found analytically \cite{roszak19a} and for $N$ nuclei is given by
	\begin{equation}
	 \rho_{01}(t) =\frac{1}{2}
	 \left(\langle\hat{\sigma}_x\rangle-i\langle\hat{\sigma}_y\rangle\right)
	 = \frac{1}{2}e^{-i(\Delta -\gamma_{e}B_{z})t}\prod_{k=0}^{N-1}L_k(t),
	    \label{rho}
	\end{equation}
	where the phase oscillation is the result of the free qubit evolution.
	The decoherence stemming from the interaction
	with the $k$-th nucleus is quantified by
	        \begin{eqnarray}
	        \label{L}
	            L_k(t) &=& \left(a_k\sin\frac{\omega t}{2}\sin\frac{\omega_{k}t}{2} + \cos\frac{\omega t}{2}\cos\frac{\omega_{k}t}{2}\right) \\
	            \nonumber
	            &&+ ip_k\left(a_k\cos\frac{\omega t}{2}\sin\frac{\omega_{k}t}{2}-\sin\frac{\omega t}{2}\cos\frac{\omega_{k}t}{2}\right).
	        \end{eqnarray}
	Here $\omega = \gamma_nB_z$ is a frequency which is determined by the applied magnetic field
	and is both typically well-known and can be controlled.
	All other parameters depend on the specific locations of the spinful carbon isotopes 
	in a given sample realization with respect to the qubit location. The $k$-dependent frequency
	(which is also magnetic-field-dependent) is
	given by
	\begin{equation}
	\label{wk}
	\omega_{k} = \sqrt{\left(\mathbb{A}_k^{z, x}\right)^2 + \left(\mathbb{A}_k^{z, y}\right)^2 + \left(\omega + \mathbb{A}_k^{z,z}\right)^2},
	\end{equation}
	and the amplitude is
	\begin{equation}
	\label{ak}
	a_k=\frac{\omega+\mathbb{A}_k^{z,z}}{\omega_{k}}.
	\end{equation}
	These parameters vary
	depending on the specific system under study and are usually not known.
	
    \section{Discrete qubit evolution \label{sec2}}

	It is evident from Eq.~(\ref{L}), where the function $L_k(t)$ determines the effect of a single environmental spin on the evolution of the qubit coherence that the information about the initial polarization of spin $k$ given by $p_k$ must manifest itself in the evolution. 
	For an environment consisting of a single spin, the imaginary part of the coherence
	(when the free evolution of the qubit is not taken into account), $\langle\hat{\sigma}_y\rangle$,
	is directly proportional to the magnitude of initial polarization, yet its evolution
	also depends on quantities which are unknown and depend on the location of the spinful nucleus
	in the diamond lattice. For environments that contain more spins, the situation becomes more
	complex, as the qubit decoherence (\ref{rho}) is proportional to a product of single-nuclear-spin
	functions (\ref{L}), so the distinction between real and imaginary components is lost while the number of free parameters grows rapidly with the growing size of the environment.	
	Nevertheless, the information about initial environmental polarization is there
	and it should be possible to extract it by studying qubit decoherence curves.
	
	In the following, we propose a straightforward way of extracting the upper bound
	on the geometric mean value of initial nuclear polarizations. To this end, we note that all
	of the terms in Eq.~(\ref{L}) have a simple cosinusoidal or sinusoidal dependence on 
	the frequency $\omega$ determined fully by the applied magnetic field.
	Since this frequency can be controlled and is known, the measurement of the qubit decoherence
	function (\ref{rho}) can be confined to set time-intervals that correspond to $\sin\frac{\omega t}{2}=0$ or $\cos\frac{\omega t}{2}=0$. We will focus on both of these cases in our further analysis.
	
	\subsection{$\cos\frac{\omega t'}{2}=0$\label{sec2a}}

        \begin{table}[]
            \centering
        \begin{ruledtabular}
            \begin{tabular}{c c c c c}
                 $k$&$r_k$~(nm)&$\mathbb{A}^{z,x}_k$~(1/$\mu$s)&$\mathbb{A}^{z,y}_k$~(1/$\mu$s)&$\mathbb{A}^{z,z}_k$~(1/$\mu$s)  \\\hline
                 $1$&2.16954&-0.017784& -0.007206& -0.034589\\
			$2$&2.33189& 0.013514 &-0.003004 &-0.028353\\
			$3$&1.03132 &0.505446 &-0.135434 &0.257915\\
			$4$&2.45435 &-0.001277 &0.004767 &-0.025882\\
			$5$&2.55104 &0.017444 &-0.028528 &0.001358\\
			$6$&1.54291 &-0.137190 &0.036760 &0.122291\\
			$7$&1.03132 &-0.251613 &-0.251613 &-0.215682\\
			$8$&2.35773 &-0.041094& -0.011011 &0.027203\\
            \end{tabular}
            \caption{Coupling constants between the NV-center qubit and nuclear spins 
			for the randomly generated realization of the environment used in 
			all figures of the main text. The second column contains distances between the qubit and each nucleus.}
            \label{table1}
        \end{ruledtabular}
        \end{table}

	\begin{figure}[!tb]
		\centering
		\includegraphics[width=1.\columnwidth]{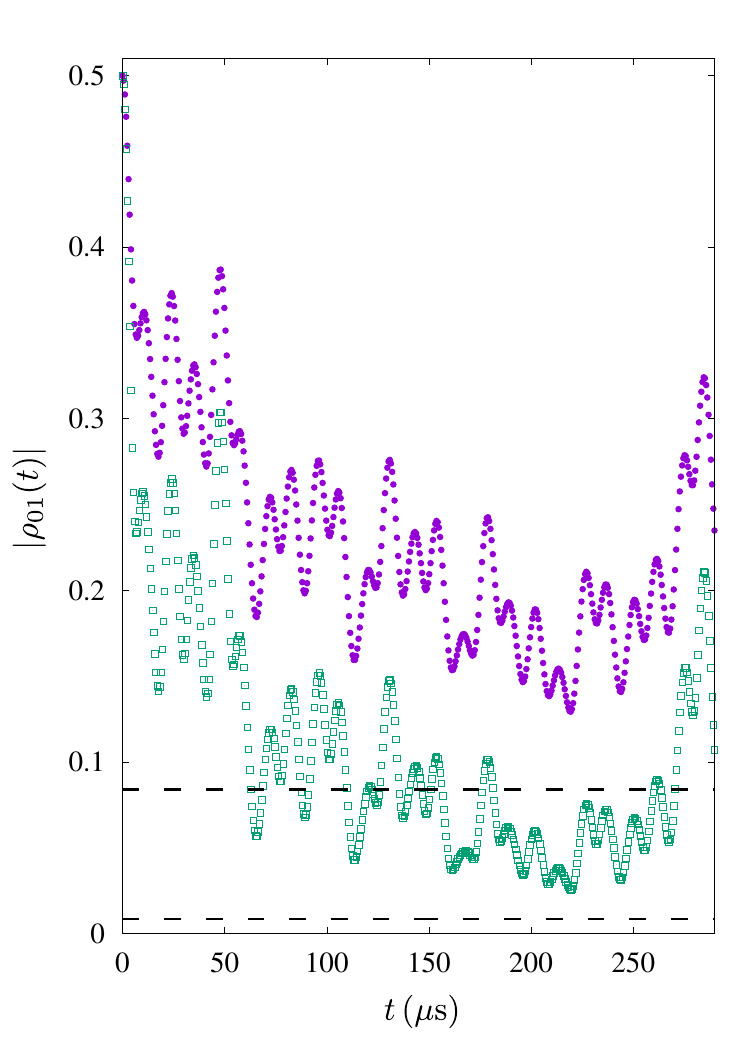}
		\caption{Absolute value of the discrete qubit coherence evolution at times $t'$ under the influence of an 8-nucleus environment and applied
			magnetic field of $B_z=1$ T. Violet dots correspond to all $p_k=0.8$ and green squares to $p_k=0.6$.
			The values of coherence corresponding to those polarizations ($8.39\times 10^{-2}$ for $p_k=0.8$ and $8.40\times 10^{-3}$ for $p_k=0.6$) 
			are marked by the dashed black lines. 
		}\label{fig1}
	\end{figure}

\begin{figure}[!tb]
\centering
\includegraphics[width=1.\columnwidth]{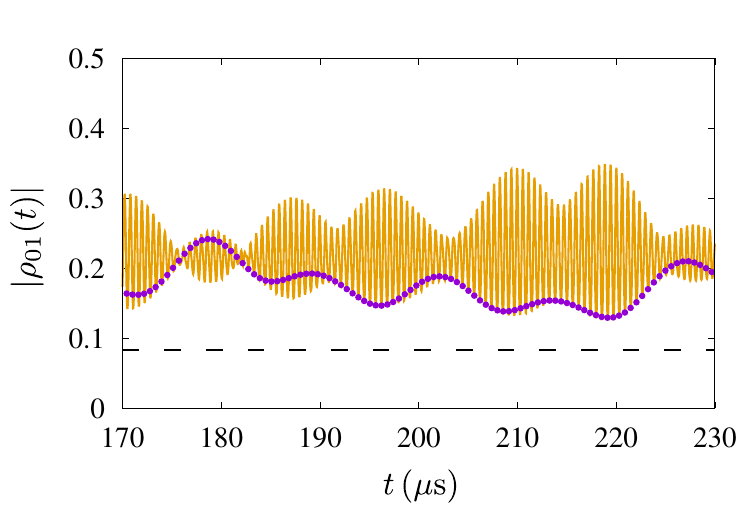}
\caption{A close-up of the $p_k=0.8$ discrete qubit coherence evolution of Fig.~\ref{fig1} (violet dots).
	The yellow curve shows the continuous line obtained from eq.~(\ref{modrho}), which contains the information
	about the interaction and the initial polarization of the environment, but is still vastly simplified
	with respect to the full qubit decoherence curve.
}\label{fig2}
\end{figure}
	
	Let us first study the situation when the value of the qubit coherence is probed only at 
	times $t'=\frac{2\pi}{\omega}(n+\frac{1}{2})$, where $n\in\mathbb{N}$. This situation corresponds to $\cos\frac{\omega t'}{2}=0$. This 
	significantly simplifies the contribution of an individual nuclear spin (\ref{L}) yielding
	\begin{equation}
	\label{l1}
	L_k(t') = \pm a_k\sin\frac{\omega_{k}t'}{2} \mp ip_k\cos\frac{\omega_{k}t'}{2}.
	\end{equation}
	It is relevant to note here that the amplitude $a_k$ given by Eq.~(\ref{ak}) is real
	and $|a_k|\in[0,1]$, hence
	\begin{equation}
	\label{l1mod}
	|L_k(t')|=\left[(a_k^2-p_k^2)\sin^2\frac{\omega_{k}t'}{2}+p_k^2\right]^{\frac{1}{2}}.
	\end{equation}
	This means that $|L_k(t)|$ oscillates between $|a_k|$ and $|p_k|$ with the frequency 
	$\omega_{k}$, while whether $|a_k|$ or $|p_k|$ is the maximum or the minimum depends on the actual
	values of the polarization of a given nuclear spin and the location-dependent amplitude. 
	
	As a typical environment of an NV-center spin qubit involves more than one spin, the bounds  
	on the oscillation of 
	\begin{equation}
	\label{modrho}
	|\rho_{01}(t')| =\frac{1}{2}\prod_{k=0}^{N-1}|L_k(t')|
	\end{equation}
	cannot be used to extract the initial polarization of the environment unless one can ensure
	that the coefficients $|a_k|\approx 1$ for all $k$. This can be achieved when [as seen in Eq.~(\ref{ak})] the applied magnetic field
	dependent frequency $\omega$ is large
	in comparison with the coupling constants of Eq.~(\ref{duzea}). We will comment on the actual
	magnitude of the necessary magnetic field later, when we present results for a realistically
	modeled NV-center spin qubit. 
	For a sufficiently large magnetic field we have $|L_k(t')|\in\left[|p_k|,1\right]$ for all $k$,
	so the coherence at the discrete time-instants $t'$ (\ref{modrho}) is bounded from below
	by the product of the environmental spin polarizations,
	\begin{equation}
	\label{modrho2}
	|\rho_{01}(t')| \in\frac{1}{2}\left[\prod_{k=0}^{N-1}|p_k|,1\right].
	\end{equation}
	Hence, the minimum value of the coherence which can be measured constitutes an upper bound
	on the product of the polarizations of the relevant environmental spins. Note here that the time-dependence of $|\rho_{01}(t')|$ involves a product of $N$ oscillatory functions,
	with different frequencies dependent on the position of spin $k$. This means that obtaining a 
	good approximation of the product of environmental spin polarizations will require measurement
	of the coherence (at discrete time instants) for long times. We will again comment on the actual times
	necessary when presenting exemplary evolutions and approximated polarizations below. 
	
	The coherence stemming from the interaction with the single carbon nucleus oscillates between its
		polarization and some magnetic field-dependent parameter, which also depends on the unknown
		NV-nuclear spin coupling constants. For a sufficiently large magnetic field (compared to the
		coupling constants) the value of this parameter approaches unity, and then the minimum of the
		coherence depends only on the nuclear spin polarization. Naturally, this estimation works better for
		weakly coupled nuclear spins, since the required magnetic field would be smaller. Even for a
		strongly interacting environment this method may give the same polarization value after reaching a
		specific value of the magnetic field. In general, the tightness of the bound can be arbitrarily small,
		and the only limitation is the applied magnetic field.
	
	\begin{figure}[!tb]
		\centering
		\includegraphics[width=1.\columnwidth]{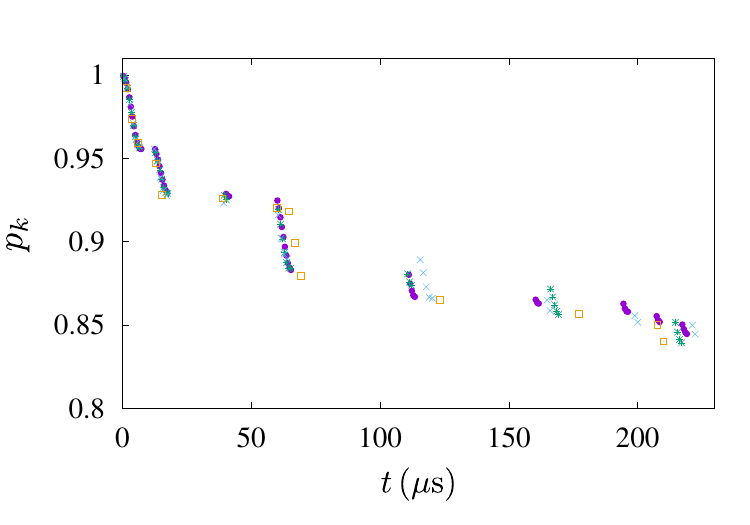}
		\caption{Prediction of average initial polarization as a function of time for $p_k=0.8$ for different
			values of the magnetic field: $B_z=1$~T - violet points; $B_z=0.75$~T - green stars; $B_z=5$~T - blue crosses; $B_z=0.25$~T - yellow squares. Points are marked
			on the plot only when the discrete qubit evolution signified a new (lower) value than previously.
		}\label{fig3}
	\end{figure}

	We demonstrate the operation of the scheme using an environment of $8$ spins at randomly generated locations
	for a qubit initially in an equal superposition state.
	The values of the coupling constants present in Eq.~(\ref{duzea}) (as well as their distance from the NV center) are given in Table \ref{table1}. They were calculated using the realistic model of the diamond crystal lattice with randomly generated spinful ${}^{13}C$ atoms. Fig.~\ref{fig1} contains the discrete evolution of the qubit coherence for all polarizations $p_k=p=0.8$
	(violet dots) and $p_k=p=0.6$ (green squares) at magnetic field $B_z=1$ T. 
	All of the initial polarization values were chosen to be equal only for convenience.
	The method yields an upper bound on the product of nuclear polarizations, and does not require
	them to be the same.
	Such a value of the magnetic field is large for the given setup in the sense that all of the amplitudes 
	$a_k\ge 0.997$.
	The values of the coherence which correspond to the actual 
	product of initial nuclear polarizations ($0.5p^8$ for this realization) are marked by the horizontal dashed lines.
	Due to the large number of nuclear spins, even the simplified function for the qubit decoherence,
	which is the outcome of the specific choice of measurement times is quite complex. This can be clearly
	seen on Fig.~\ref{fig2} which contains a close-up of the discrete evolution of qubit coherence and the 
	corresponding continuous curve given by Eqs. (\ref{l1mod}) and (\ref{modrho} for $p=0.8$. 
	
	More figures analogous to Fig.~\ref{fig1} for a different distribution of environmental spins can be 
	found in the appendix.
	
	Fig.~\ref{fig3} contains a plot where the lowest predicted value of the average single-nucleus polarization, 
	$\overline{p}=\sqrt[N]{\prod_{k=0}^{N-1}p_k}$,
	is marked as a function of the first time that this value was reached. Since in the studied case we chose all $p_k=p$, we obviously have $\overline{p}=p$,
	and the plot is made for $p=0.8$.
	The different types of points correspond
	to different values of the magnetic field, with the lowest value being $B_z=0.25$~T, which yields all amplitudes 
	$a_k\ge 0.975$ for this realization ensuring the capability of detection of the product of initial nuclear polarization if the smallest $p_k<0.975$. As can be seen, the minimum value obtained during the discrete 
	measurements of the evolution initially changes often, up to times of a few hundred $\mu$s.
	Further change occurs at around $1600$~$\mu$s and is not included in the plot, since the measurement
	of such coherence times would be unrealistic taking into account the longest experimentally reported
coherence times in such systems which are of similar order \cite{herbschleb19}. 
	
	We find that the functions which are responsible for the discrete evolution of the qubit are strongly 
	and qualitatively
	influenced by the number of environmental spins and the actual values of the coupling constants (the distribution
	of the nuclear spins), while the initial polarization has a mostly quantitative effect. If the magnetic field is 
	high enough to guarantee that the amplitudes $|a_k|$ are close to unity, then also its effect on the the
	operation of the scheme is negligible as seen in Fig.~\ref{fig3}.

\subsection{$\sin\frac{\omega t''}{2}=0$\label{sec2b}}

	Predictably, the ``opposite'' situation when the qubit coherence is only probed at times $t''=\frac{2\pi}{\omega}n$ when 
	$\sin\frac{\omega t''}{2}=0$ is also of interest for the approximation of initial environmental
	polarization. In this situation, the contribution of the individual nuclear spin is given 
	by 
	\begin{equation}
	\label{2}
	L_k(t'') = \pm\cos\frac{\omega_{k}t''}{2}\pm ip_ka_k\sin\frac{\omega_{k}t''}{2}.
	\end{equation}
	This yields the absolute value
	\begin{equation}
	\label{l2mod}
	|L_k(t'')|=\left[(1-a_k^2p_k^2)\cos^2\frac{\omega_{k}t''}{2}+a_k^2p_k^2\right]^{\frac{1}{2}},
	\end{equation}
	with unambiguously defined upper and lower bounds $|L_k(t'')|\in[|a_k||p_k|,1]$.
	This means that the approximation of the upper bound on $\prod_{k=0}^{N-1}|a_k||p_k|$
	does not require any additional assumptions (such as a high magnetic field required in 
	Subsection \ref{sec2a}), since
	\begin{equation}
	\label{modrho3}
	|\rho_{01}(t'')| \in\frac{1}{2}\left[\prod_{k=0}^{N-1}|a_k||p_k|,1\right].
	\end{equation}
	
	Nevertheless to obtain information about the initial polarizations not weighted by the 
	amplitudes $|a_k|$, knowledge of the product of amplitudes $\prod_{k=0}^{N-1}|a_k|$ is required. 
	This problem can
	be solved in two ways. The first, analogous to the solution in Subsection \ref{sec2a} requires
	a high magnetic field to be applied, which results in $\prod_{k=0}^{N-1}|a_k|\approx 1$
	and consequently reduces the lower bound on the coherence at time-instants $t''$
	to $\frac{1}{2}\prod_{k=0}^{N-1}|p_k|$. Results obtained in this regime are very similar to the results
	of the previous subsection, so we do not include additional plots.
	The relevant difference between this case and the results of Sec.~\ref{sec2b} is the necessary value of the magnetic field. Previously
	the magnetic field had to be high enough that the amplitude $|a_k|$ for each individual nucleus would be
	greater than the polarization $p_k$ of this nucleus, while here only the value of the product of all amplitudes is of importance.
	
	The other solution requires a measurement of $\prod_{k=0}^{N-1}a_k$, which is possible 
	using the method of Sec.~\ref{sec2a}. Note that at times $t'$, if the environment has 
	not been polarized, so all $p_k=0$, the qubit coherence is bounded as
	\begin{equation}
	\label{modrhoa}
	|\rho_{01}(t')| \in\frac{1}{2}\left[0,\prod_{k=0}^{N-1}|a_k|\right],
	\end{equation}
	regardless of the magnetic field, as long as the frequency $\omega$ does not cancel out with any coupling constant $\mathbb{A}_{k}^{z,z}$ for some nuclear spin (in which situation a small change of the applied field rectifies 
	the situation). Hence, the maximum value of $|\rho_{01}(t')|$
	obtained during the evolution of the coherence yields the lower bound on $\prod_{k=0}^{N-1}|a_k|$
	and can be used for the estimation of the product of initial environmental polarizations
	from Eq.~(\ref{modrho3}), as long as both measurements are performed at the same magnetic field
	(since $\prod_{k=0}^{N-1}|a_k|$ depends strongly on the magnetic field). This second solution 
	has the advantage that it does not require high magnetic fields to ensure the accuracy of the estimation
	of the initial polarization, but it does require a second measurement of qubit decoherence to
	be performed for an unpolarized environment, and this measurement yields only a lower 
	bound on the actual value of the product of amplitudes.

	\section{Conclusion \label{sec6}}
	
		We have proposed a method for the estimation of the product of initial polarizations of the spinful
		nuclei of the ${}^{13}C$ carbon isotope in diamond via the measurement of the coherence of a single NV-center
		spin qubit which evolves due to the interaction with such an environment. 
		Taking an ensembe of NV center spins into account would mean that the result
			would be averaged over different decoherence that would be demonstrated by each spin qubit (since
			the locations of each NV center relative to the spin environment would differ form qubit to qubit,
			yielding a different decoherence curve every time), making the idea of generalizing the method
			to an ensamble of NV centers unfeasible. On the other hand, it is also unnecessary, since single NV
			center systems exist and are used in e.~g.~magnetometry.
		
		We have found, that although
		the amount of unknown variables describing the interaction in physical realizations 
		and the complexity of the function 
		that governs the decoherence due to the presence of already a small number of carbon isotope atoms,
		precludes a straightforward readout of the polarization, measurement of the coherence only at some 
		points of time allows to obtain much more information. The points of time are chosen in such a way
		as to significantly simplify the functions which are responsible for the influence of each environmental
		spin on the qubit and they depend only on the applied magnetic field. The minimum of the decoherence 
		curves obtained in such a way are directly proportional to the upper bound on the product of the initial
		environmental polarizations (the proportionality factor is the initial qubit coherence).
		
		We exemplified the operation of the scheme on an environment of eight nuclear spins distributed randomly
		in the vicinity of the qubit in the diamond crystal lattice. Due to the nature of the short-range interaction between the qubit and its environment, the number of nuclear spins giving a noticeable contribution will not be significantly larger in the experimental setup.
		The method yields an upper bound on the actual value of the product of polarizations,
		and procures a rough estimate of the value at longer times.
		In the example used we have obtained the average measured polarization of $0.84463$ for the situation
		when all $p_k=0.8$. This was obtained at measurement time $t=219$ $\mu$s.
	
		The biggest advantage of the proposed method is the simplicity of the required measurements, which do not involve accessing the degrees of freedom of the environment in any way. In fact, all of
		the information about the state of the environment is gained due to information being transferred
	into the qubit state during the joint system-environment evolution. Since this process is 
		only facilitated by the interaction between the NV-center spin qubit and it's environment of nuclear spins,
		also no operations on the qubit are required, except for well timed measurement of qubit the coherence.

\appendix
\section{Discrete qubit coherence evolution}\label{sec:Appendix}

Here, we present additional figures of the absolute value of the discrete qubit coherence evolution at times $t'$ (in analogy to Fig. \ref{fig1}) given by the Eq. \ref{modrho}, together tables of different coupling constants $\mathbb{A}^{z,i}_{k}$ used to generate them. 
	Each table corresponds to the indicated figure. All figures contain the discrete evolution of the qubit coherence for equal initial polarizations $p_k = p = 0.8$ (violet dots) and $p_k = p = 0.6$ (green squares) at magnetic field $B_z = 3~T$. The values of the coherence which correspond to the actual product of initial nuclear polarizations ($0.5p^{8}$ for this realization) are marked by the horizontal dashed lines.

The number of environmental spins taken into account for each plot remains $8$ as in Fig.~\ref{fig1},
but the random distribution is different for each one.
This is to keep the figures easily comparable to Fig.~\ref{fig1}, as the purpose of the figures is to 
exemplify the range of different evolutions that can be induced by the interaction of the NV-center with
comparable spin environments.

\begin{table}[htp]
            \centering
        \begin{ruledtabular}
            \begin{tabular}{c c c c c}
                 $k$&$r_k$~(nm)&$\mathbb{A}^{z,x}_k$~(1/$\mu$s)&$\mathbb{A}^{z,y}_k$~(1/$\mu$s)&$\mathbb{A}^{z,z}_k$~(1/$\mu$s)  \\\hline
            $1$&0.50442 &1.37617   & 0        &0.973096\\
			$2$&0.56396 &-0.689293 & 0.170556 &-0.417774\\
			$3$&0.61778 & 0.499393 & 0        &-0.353124\\
			$4$&0.63680 &-0.013411 &-0.116145 &-0.47416\\
            $5$&0.68492 &-0.372671 & 0.161371 &-0.22399\\
            $6$&1.03132 &-0.251613 &-0.251613 &-0.215682\\
            $7$&1.54291 &-0.137190 &0.036760  &0.122291\\
			$8$&2.55104 & 0.017444 &-0.028528 &0.001358\\
            \end{tabular}
            \caption{Coupling constants 
			(as in Table {\ref{table1}}) used to generate Fig.~(\ref{fig4})}
            \label{table2}
        \end{ruledtabular}
        \end{table}

        \begin{figure}[htp]
		\centering
		\includegraphics[width=0.95\columnwidth]{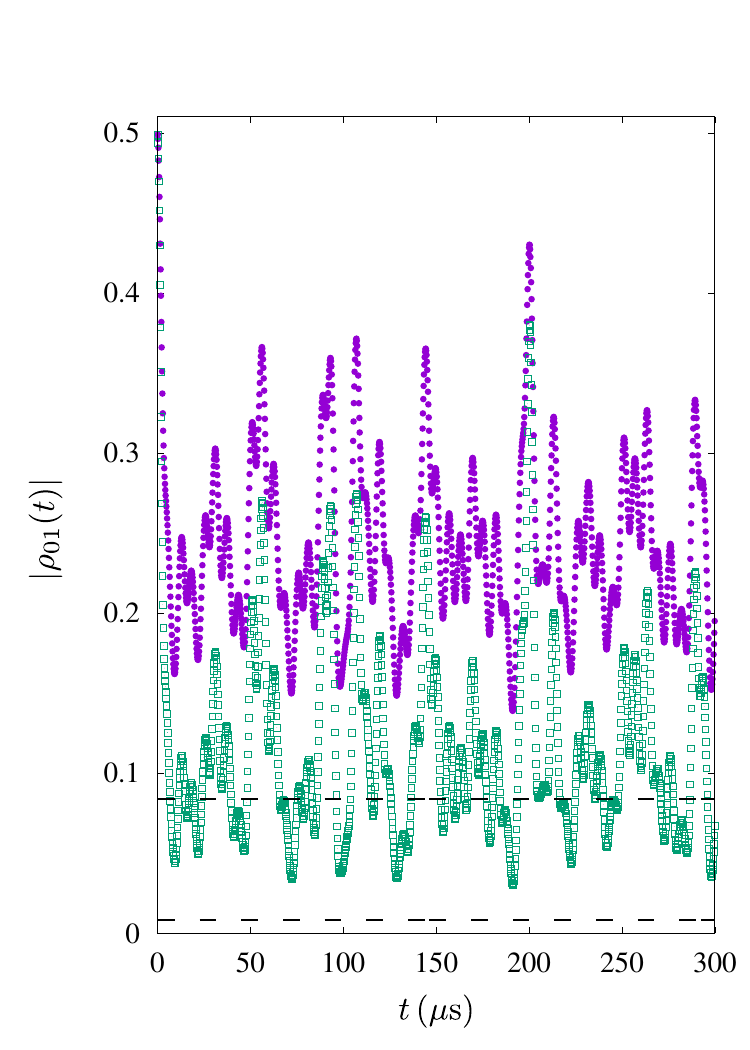}
		\caption{Absolute value of the discrete qubit coherence evolution (as Fig.~\ref{fig1})for coupling constants taken from Table \ref{table2}.
		}\label{fig4}
	\end{figure}

\begin{table}[htp]
            \centering
        \begin{ruledtabular}
            \begin{tabular}{c c c c c}
                 $k$&$r_k$~(nm)&$\mathbb{A}^{z,x}_k$~(1/$\mu$s)&$\mathbb{A}^{z,y}_k$~(1/$\mu$s)&$\mathbb{A}^{z,z}_k$~(1/$\mu$s)  \\\hline
            $1$&0.56396 & 0.196941 & 0.682223 &-0.417774\\
            $2$&0.56396 & 0.492352 &-0.511667 &-0.417774\\
            $3$&0.66728 & 0        & 0        &-0.420338\\
            $4$&1.03132 &-0.251613 &-0.251613 &-0.215682\\
            $5$&1.54291 &-0.137190 & 0.036760 & 0.122291\\
            $6$&2.16954 &-0.017784 &-0.007206 &-0.034589\\
            $7$&2.35773 &-0.041094 &-0.011011 &0.027203\\
			$8$&2.55104 & 0.017444 &-0.028528 &0.001358\\
            \end{tabular}
        \caption{Coupling constants 
        (as in Table {\ref{table1}}) used to generate Fig.~(\ref{fig5})}
            \label{table3}
        \end{ruledtabular}
        \end{table}

        \begin{figure}[htp]
		\centering
		\includegraphics[width=0.95\columnwidth]{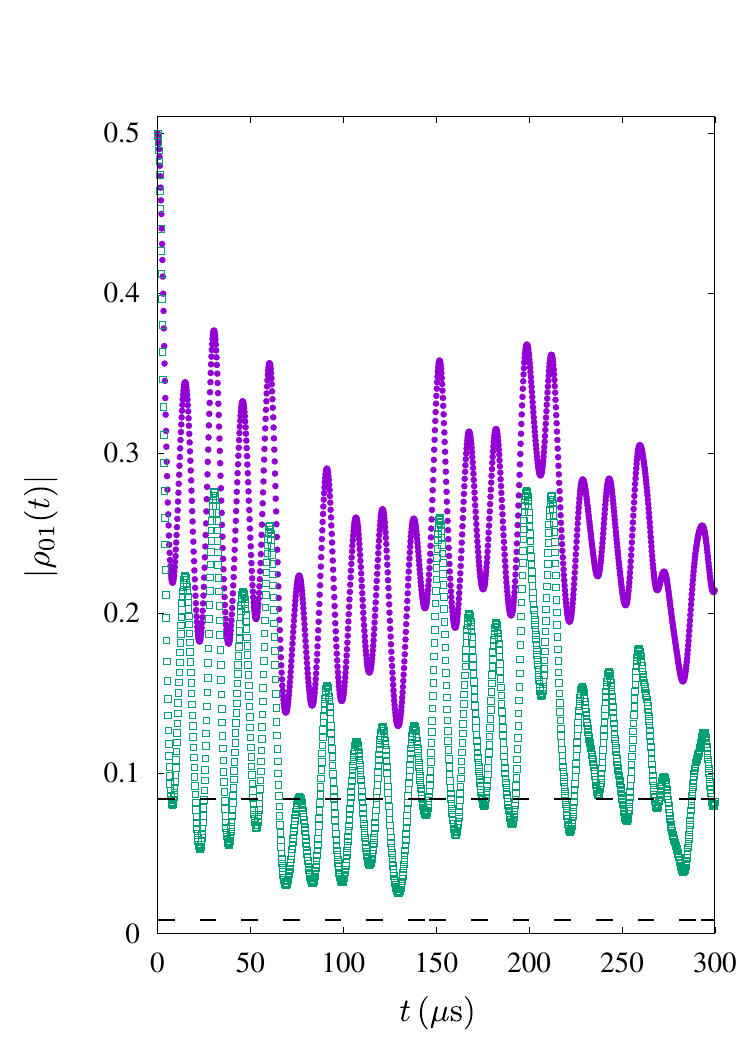}
		\caption{Absolute value of the discrete qubit coherence evolution (as Fig.~\ref{fig1})for coupling constants taken from Table \ref{table3}.
		}\label{fig5}
	\end{figure}

\begin{table}[htp]
            \centering
        \begin{ruledtabular}
            \begin{tabular}{c c c c c}
                 $k$&$r_k$~(nm)&$\mathbb{A}^{z,x}_k$~(1/$\mu$s)&$\mathbb{A}^{z,y}_k$~(1/$\mu$s)&$\mathbb{A}^{z,z}_k$~(1/$\mu$s)  \\\hline
            $1$&0.50442 &1.37617   & 0        & 0.973096\\
            $2$&0.56396 & 0.492352 &-0.511667 &-0.417774\\
            $3$&0.61778 & 0.499393 & 0        &-0.353124\\
            $4$&0.63680 &-0.013411 &-0.116145 &-0.47416\\
            $5$&0.66728 & 0        & 0        &-0.420338\\
            $6$&1.03132 &-0.251613 &-0.251613 &-0.215682\\
            $7$&1.54291 &-0.137190 & 0.036760 & 0.122291\\
            $8$&2.16954 &-0.017784 &-0.007206 &-0.034589\\
            \end{tabular}
        \caption{Coupling constants 
        (as in Table {\ref{table1}}) used to generate Fig.~(\ref{fig5})}
            \label{table4}
        \end{ruledtabular}
        \end{table}

        \begin{figure}[H]
		\centering
		\includegraphics[width=0.95\columnwidth]{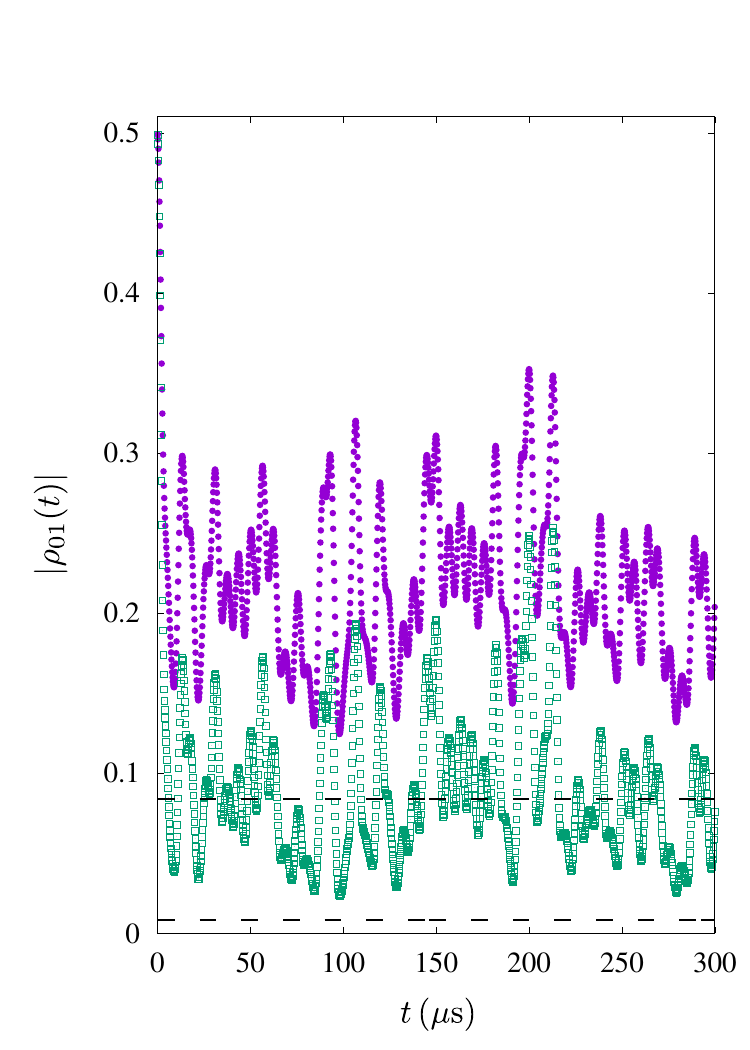}
		\caption{Absolute value of the discrete qubit coherence evolution (as Fig.~\ref{fig1})for coupling constants taken from Table \ref{table4}.
		}\label{fig6}
	\end{figure}

\end{document}